\documentclass[selectedoptions]{aipproc}

\def \sss{\scriptscriptstyle}
\def \e{\vec E}
\def \b{\vec B}
\def \r{R}
\def \k{\hat \eta}
\def \fp{\hat \eta^{\sss FP}}
\def \thb{\theta_B}
\def \bd{\beta_{\sss D}}
\def \gd{\gamma_{\sss D}}
\def\lesssim{\hbox{\hspace{1.5mm}}\raise2pt
       \vbox{\hbox{$<$}}\lower2pt
       \vbox{\moveleft9.0pt\hbox{$\sim$ }}\hbox{\hskip 0.05mm}}

\layoutstyle{6x9}
\title[Rotationally-induced asymmetry]{Rotationally-induced 
      asymmetry\\ in the double-peak lightcurves\\ of the bright EGRET pulsars?} 

\author{J. Dyks}{
 address={Nicolaus Copernicus Astronomical Center, Toru{\'n}, Poland}
 }
\author{B. Rudak}{
 address={Nicolaus Copernicus Astronomical Center, also
          TCfA NCU, Toru{\'n}, Poland}
}

\begin{abstract}
Pulsed emission from the bright EGRET pulsars  - Vela, Crab, and Geminga 
- extends up to $\lesssim 10$ GeV.
The generic gamma lightcurve features two peaks separated by 0.4 to 0.5 
in phase. According to Thompson (2001) 
the lightcurve becomes asymmetrical above $\sim 5$~GeV in such a way 
that the trailing peak dominates over the leading peak.
We attempt to interpret this asymmetry within a single-polar-cap scenario.
We investigate the role of rotational effects on the magnetic 
one-photon absorption rate in inducing such asymmetry.
Our Monte Carlo simulations of pulsar gamma-ray beams reveal that 
in the case of oblique rotators with rotation periods
of a few millisecond the rotational effects lead to the asymmetry 
of the requested magnitude.
However, the rotators relevant for the bright EGRET pulsars 
must not have their inclination angles too large in order to keep 
the two peaks at a separation of $\sim 0.4$ in phase. 
With such a condition imposed on the model rotators
the resulting effects are rather minute and can hardly be reconciled 
with the magnitude of the observed asymmetry.
\end{abstract}

\begin{document}
\maketitle

\section{Introduction}
High quality gamma-ray data 
for three pulsars - Vela, Crab, and Geminga - provided by EGRET aboard the CGRO enable
an analysis of properties of pulsar high-energy radiation as a function of both, photon energy and
phase of rotation.
The spectra of pulsed radiation from these sources (as well as from three other EGRET pulsars:
B1706-44, B1951+32, and B1055-52) 
extend up to $\lesssim 10$~GeV. 
All three pulsars feature gamma lightcurves characterised by two strong peaks separated
by 0.4 to 0.5 in rotational phase.
The pulses are asymmetrical in the sense that their leading peaks (LP) exhibit lower
energy cutoffs (about $\sim 5$~GeV) than their trailing peaks (TP). In other words, the trailing peaks 
dominate over
the leading peaks above $\sim 5$~GeV \cite{t2001}.

The high-energy cutoffs in pulsar spectra
are interpreted within polar cap models as due to one-photon absorption of gamma-rays in strong
magnetic field with subsequent $e^\pm$-pair creation.
A piece of observational support for such an interpretation comes from   
a strong correlation between the inferred `spin-down' magnetic field 
strength and the position of the high-energy cutoff \cite{t2001}.
This, in turn, opens a possibility that the observed asymmetry between LP and TP, i.e.
the dominance of LP over TP above $\sim 5$~GeV,
is a direct consequence of propagation effects (which eventually lead to stronger one-photon absorption 
for photons forming LP than TP)
rather than due to some inherent property of the gamma-ray emission region itself.
The aim of this research note is to investigate the role of rotation in the built-up
of such asymmetry.

We consider the purely rotational effects: due to the presence of the rotation-induced electric
field $\e$, aberration of photon direction and 
slippage of magnetosphere under the photon's path.
All these effects are of the same order of magnitude: $\propto \beta$, where $\beta$ is a local
corotation velocity $v$ expressed in units of the speed of light $c$.
We assume that the magnetic field within the radius of a few pulsar radii has a shape of a rigidly rotating
static-like dipole.
In reality the rotation distorts this shape
because of both retardation effects as well as toroidal currents 
due to plasma dragging. Fortunately, these are 
higher order effects 
($\propto \beta^2$) and will be ignored below.
Another expected disturbance of the magnetic field structure comes from longitudinal
currents suspected to flow within the region of open field lines.
No self-consistent solution of this problem 
has been found so far (see \cite{besk} and references therein). 
Nevertheless, the longitudinal currents are expected to modify the dipole magnetic field by a factor 
of $\propto \beta^{3/2}$,
and therefore they will be neglected too.

\section{Is the leading peak absorbed more efficiently?}

\begin{figure}
\resizebox{\textwidth}{!}
{\includegraphics{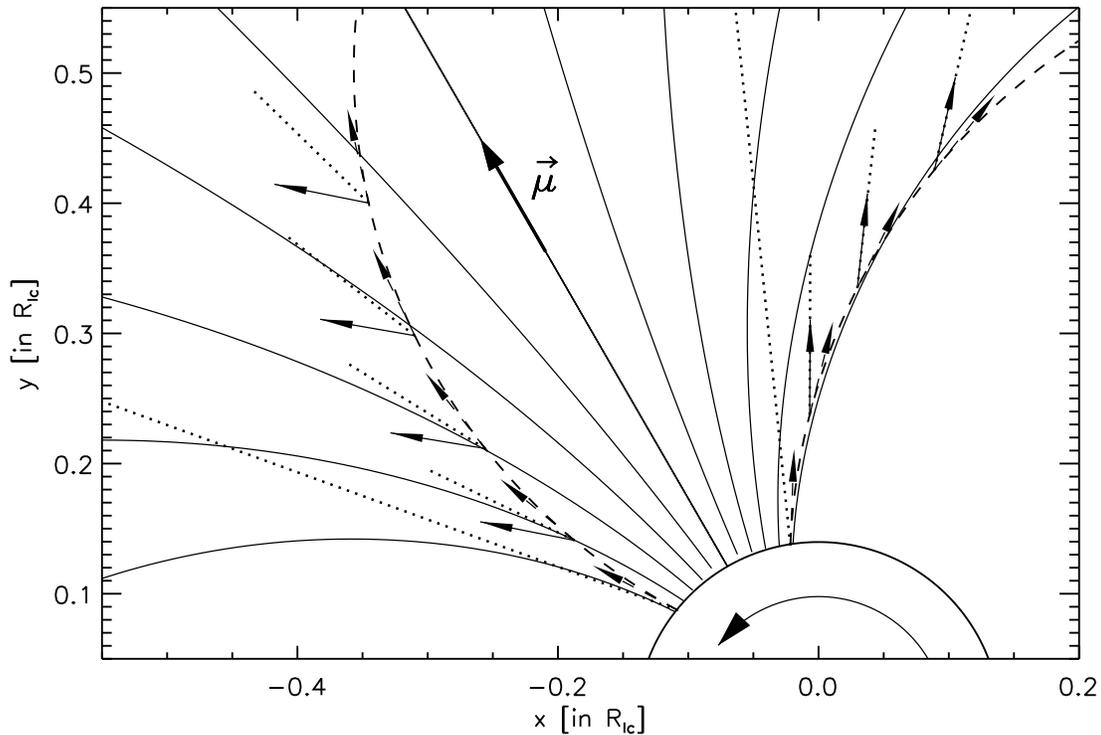}}
\caption{The view of the orthogonally rotating pulsar with the period
$P=1.5$ ms. The magnetic field lines correspond to the static-like dipole
(see text for other details).}
\label{fig1}
\end{figure}

Let us consider the photon propagation in the equatorial 
plane of an orthogonal rotator with rotation period $P=1.5$ ms (this case is ideal for 
instructive purposes, and it is shown in Fig~1).
For definitness, we
constrain to its northern magnetic hemisphere.
Then, at any point in the plane, photons propagating upwards cross 
the magnetic field $\b$ at angles $\thb < 90^\circ$
and the rotation-induced electric field $\e=-\vec \beta \times \b$
is directed towards the reader, at right angle to the page.

One of the striking features of one-photon magnetic pair production 
is that its rate $\r$ is not axisymmetric around a local $\b$ direction
if a weak electric field $\e \perp \b$ is present
\cite{dl75}. Within the accesible range of angles $\thb$
the rate vanishes in the unique direction which lies in the plane
perpendicular to $\e$ and deviates from $\b$ by angle
$\sim E/B$ towards the rotation direction.  
In local coordinate frame with $\hat z \parallel \b$, 
$\hat y \parallel \e$, and $\hat x \parallel \e\times\b$ this
"free propagation" direction is given by
$\fp = [\eta^{\sss FP}_x, \eta^{\sss FP}_y, \eta^{\sss FP}_z] = [E/B, 0, (1 - E^2/B^2)^{1/2}]$.
If $E\ll B$ the pair production rate $\r$
is approximately symmetric around the free propagation direction $\fp$
instead of around $\b$. 
Moreover, $\r$ increases monotonically for angles which depart 
from $\fp$.
Accordingly, the projection of photon momentum 
on $\fp$ is a better measure of $\r$ than its projection on $\b$.
The directions of $\fp$ at various points within the magnetosphere
are shown in Fig.~1 as solid arrows.
Note that the free propagation direction deviates from the local $\b$
by an angle $\thb = \arcsin(E/B) \simeq E/B$ which increases with altitude.

High-energy photons are emitted from the outer rim of standard polar cap 
tangentially to the magnetic field in the corotating frame
(dashed arrows at the star surface in Fig.~1).
In the observer frame (OF) the photons propagate at the aberrated
direction (dotted lines) which at the emission point is just the free propagation direction
(at this point the angle between the photon direction and the magnetic field line 
equals
$\theta_B \simeq E/B$, and therefore
$\r = 0$ \cite{hte78}, \cite{zzq98}). 
Initially, therefore, the rate $\r$
is symmetric for photons in the leading and in the trailing peak
both in the corotating and in the observer frame. As the photons propagate outward,
however, the free propagation direction starts to deviate from the photon
direction $\k$ and this occurs in a different way for photons of the leading and the trailing
peak.
The reasons for which the local $\fp$ diverges from $\k$ include:
(1) the magnetic field line curvature, (2) the increase in $E/B$ ratio with
altitude, and (3) the slippage of magnetic field lines under a photon's path.
For photons in the leading peak the effects (1) and (2)
cumulate  whereas for the trailing peak they effectively tend to cancel out each other. 
In consequence, the photons in the leading peak suffer stronger absorption
than the photons of the same energy in the trailing peak. This is why
the high energy cutoff in the LP spectrum occurs at
a slightly lower energy than the cutoff for the TP.
The difference becomes more pronounced for smaller curvature of magnetic
field lines. The slippage (3) does not change this picture.

Photon propagation direction $\k$ as seen in the observer frame
(dotted lines) and the local 
free propagation direction
$\fp$ in the OF (solid arrows) are shown in Fig.~1 for 
a few positions along the photon trajectory 
in the corotating frame. The stronger absorption of
the leading peak is evident (for the TP, $\k$ nearly coincides with $\fp$).

\begin{figure}
\resizebox{\textwidth}{!}
{\includegraphics{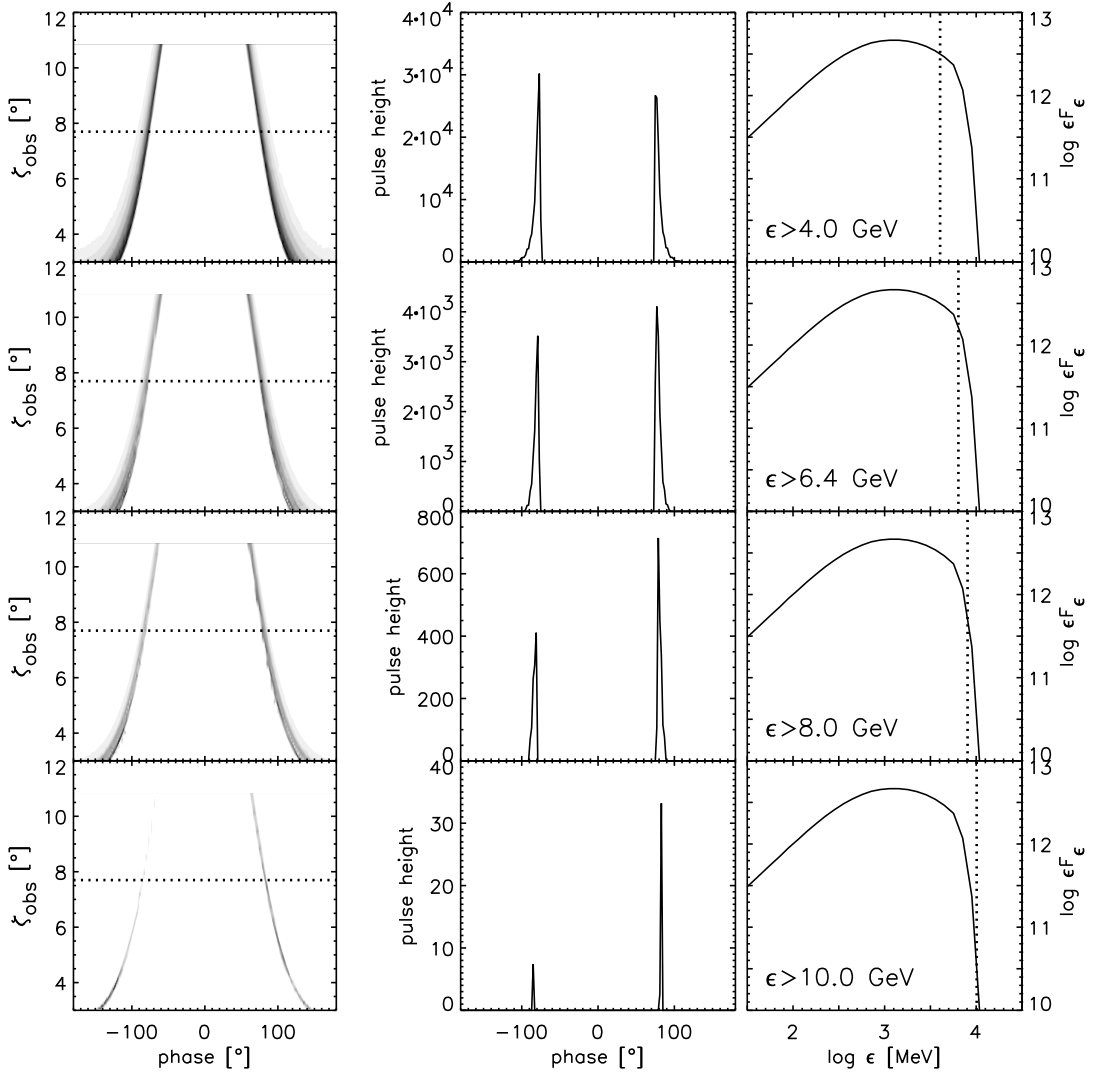}}
\caption{Evolution of double peak pulse profile with photon energy.
For details see text.}
\label{fig2}
\end{figure}

Another way to understand the asymmetry in the pair production rate is to
follow photon trajectories in a reference frame where
 $\e^\prime = 0$ is fulfilled\footnote{
This filled and corotating magnetosphere assumption is an approximation
within the polar gap where accelerating electric field is expected.
Its strength is lower  
than the corotational electric field by $\beta^{1/2}$.}
and the pair production rate $\r^\prime$ can be approximated by the well-known formula for the pure-$\b$ case:
$\r^\prime(\epsilon^\prime, \b^\prime, \sin\thb^\prime) = c_1
\sin\thb^\prime B^\prime \exp{[-c_2/ (\epsilon^\prime \sin\thb^\prime 
B^\prime)]}$
where $\thb^\prime = \angle(\k^\prime, \b^\prime)$ and
$\epsilon^\prime$ is photon energy.
If a frame of local $\e\times\b$ drift is chosen, 
with $\vec \bd
= \e\times\b/B^2$, the rate $\r$ in the observer frame can be expressed
as $\r = \gd\ (1 - \eta_x\bd)\ \r^\prime(\epsilon^\prime,
\ \b^\prime,\ \sin\thb^\prime)$ where
$\epsilon^\prime = \epsilon\gd(1 -
\eta_x\bd)$, $\b^\prime = \b/\gd$, and $\sin\thb^\prime=
[(\eta_x - \bd)^2 + \eta_y^2(1 - \bd^2)]^{1/2}/(1 - \eta_x\bd)$.
In the equatorial plane of the orthogonal rotator $\eta_y = 0$
so that $\eta_x = \mp\sin\thb$ where the signs 'minus' and 'plus'
correspond to the leading and the trailing peak, respectively.
Typically $\eta_x \ll 1$ which implies that the difference between the pair
production rates in the locally drifting frame and in the observer frame
results primarily from the aberration of photon direction whereas the change
in $\epsilon$ or $\b$ is a second order effect.
Obviously, the aberration is asymmetric for the leading and the trailing
peak ($\eta_x < 0$ and $\eta_x > 0$ respectively).
Fig.~1 presents this ``aberration effect" 
in the rigidly corotating frame where $\e^\prime=0$ is assumed. 
Photon trajectories in this frame are marked with
dashed lines which indicate clearly that photons of the leading
peak encounter larger $B^\prime_\perp$ 
than photons of the trailing peak of the same energy.

We have performed Monte Carlo simulations of radiative processes above
polar cap, including the emission of curvature radiation with
subsequent one-photon pair production (for description of the model
see \cite{dh82}).
We find that a difference by a factor of $\sim 2$
in the position of high-energy cutoffs 
for the leading and the trailing peak
can be generated 
for Vela-like objects
with emission regions placed a few stellar radii above the surface
provided the inclination angle $\alpha$ of the magnetic dipole exceedes $\sim45$ degrees.
However, for such large $\alpha$
the observed peak-to-peak separation of about $0.4$ cannot be
reproduced \cite{dr00}.
Therefore, we now turn to the case of nearly aligned rotators.
As an example we choose the model with parameters of the Vela pulsar 
($B_{\rm dip} = 3.4$ TG, $P=0.0893$ s) and we take the inclination angle
$\alpha = 7.6^\circ$. 
We then place the polar accelerator at the
altitude of $\sim 4$ R$_{\rm NS}$ to ensure that the magnetosphere is not entirely 
opaque to curvature photons of
energy $\lesssim 10$ GeV.
The numerical results are presented in Fig.~2. The three columns of Fig.~2 
present: 1) the photon distribution in the
observer's colatitude-phase space $(\zeta_{\rm obs}, \phi)$ (left),
2) pulse profiles for $\zeta_{\rm obs}=  7.7^\circ$ integrated above a photon
energy $\epsilon$ (middle), and 3) the phase-integrated spectrum with
$\epsilon$ marked with the dotted vertical line (right).
Four rows correspond to increasing energy of 
$\epsilon=4.0$, $6.4$, $8.0$, and $10$ GeV (top to bottom).
The peak-to-peak asymmetry  (middle column) 
does agree qualitatively with the observed data even
in this nearly aligned case.
However, the really strong fading of the leading peak LP  occurs only at the very 
end of the curvature spectrum, where its level drops below the detectability level of EGRET.

Therefore, the rotational effects alone probably cannot
account for the asymmetries observed
in the bright EGRET pulsars. 
However, the photon statistics at the highest energy bins
is too low to treat this conclusion as being firm.
Alternatively, the asymmetries may be generated by some 
properties inherent in the region of the gamma-ray emission.

\section{acknowledgments}
We are grateful to V.S.~Beskin for useful comments 
on the issue of magnetospheric distortions. This work was supported by KBN grant
2P03D02117 and NCU grant 405A.


\begin{thebibliography}{}


\bibitem{besk} Beskin, V.~S., 
\emph{Physics-Uspekhi}, {\bf 42}, 1071$-$1098 (1999)

\bibitem{dh82} Daugherty, J.~K., and Harding, A.~K.,
 \emph{ApJ}, {\bf 252}, 337$-$347 (1982)


\bibitem{dl75} Daugherty, J.~K., and Lerche, I.,
 \emph{ApSS}, {\bf 38}, 437$-$445 (1975)

\bibitem{dr00} Dyks, J., and Rudak, B.,
 \emph{MNRAS}, {\bf 319}, 477$-$483 (2000)

\bibitem{hte78} Harding, A.~K., Tademaru, E., Esposito, L.~W.,
 \emph{ApJ}, {\bf 225}, 226$-$236 (1978)
 


\bibitem{t2001} Thompson, D.~J., {\cal astro-ph/0101039} (2001)

\bibitem{zzq98} Zheng, Z., Zhang, B., and Qiao, G.~J.
 \emph{A\&A}, {\bf 334}, L49$-$L52 (1998)



\end{thebibliography}
\end{document}